\documentclass[12pt]{article}
\usepackage{helvet,times,mathptm}
\usepackage{graphicx}
\usepackage{color}
\usepackage{amssymb,amsmath,wasysym}
\begin{document}

\begin{center}
{\noindent
{\bf Study of Low Temperature Magnetic Properties of a Single Chain Magnet With 
Alternate Isotropic and Non-Collinear Anisotropic Units}
}
\end{center}

\begin{center}
{\small \bf Shaon Sahoo\let\thefootnote\relax\footnotetext{S. Sahoo \\ Department 
of Physics, Indian Institute of Science, Bangalore 560012, India \\ e-mail: 
shaon@physics.iisc.ernet.in \\ }, Jean-Pascal Sutter
\let\thefootnote\relax\footnotetext{ J.-P. Sutter \\  Laboratoire de Chimie de 
Coordination, Universit\'{e} de Toulouse, F-31077 Toulouse, France \\ e-mail: 
sutter@lcc-toulouse.fr \\ }and S. Ramasesha\let\thefootnote\relax\footnotetext{S. 
Ramasesha \\ Solid State $\&$ Structural Chemistry Unit, Indian Institute of Science, 
Bangalore 560012, India \\ e-mail: ramasesh@sscu.iisc.ernet.in}}
\end{center}

\begin{abstract}
\noindent
Here we study thermodynamic properties of an important class of single-chain magnets 
(SCMs), where alternate units
are isotropic and anisotropic with anisotropy axes being non-collinear. This class of
SCMs shows slow relaxation at low temperatures which results from the interplay of two 
different relaxation mechanisms, namely
dynamical and thermal. Here anisotropy is assumed to be large and                  
negative, as a result, anisotropic units behave like canted spins at low temperatures;
but even then simple Ising-type model does not capture the essential physics of the
system due to quantum mechanical nature of the isotropic units. We here show how
statistical behavior of this class of SCMs can be studied using a transfer matrix (TM)
method. We also, for the first time, discuss in detail how weak inter-chain interactions
can be treated by a TM method. The finite size effect is also discussed which becomes
important for low temperature dynamics. At the end of this paper, we apply this technique
to study a real helical chain magnet.
\end{abstract}

\noindent
{\bf Keywords} Single chain magnets \textperiodcentered Alternate isotropic and 
anisotropic units \textperiodcentered Inter-chain interaction

\section{Introduction}
In the area of magnetism, the field of single chain magnets (SCMs) has 
attracted huge interest due to its potential use in dense data storage. Magnetism 
in one dimension is not only interesting from academic point of view, it has also 
created wide interest for application in technology. A huge body of research has 
already gone into understanding the physics of SCMs with the aim of designing systems 
with desired properties \cite{bogani,miyasaka,coulon,vindigni,sun}. 

Though it is theoretically impossible to have a spontaneous magnetization at finite 
temperature in strictly one dimension, apparently 
there is no upper limit for the relaxation time of induced magnetization. In the 
field of SCMs, the focus therefore is to increase the relaxation time by combining 
two modes of relaxations, namely, dynamic relaxation (creation and movement of domain 
wall(s) along chain) and thermal relaxation (hopping across anisotropy barrier). This 
is expected to be achieved by synthesizing a spin chain with anisotropic units. While 
synthetic chemists are exploring different chemical possibilities, theoreticians are 
developing new techniques to study these compounds. The theoretical studies in return 
help to understand the paradigms involved in increasing the relaxation time. 

Modeling a SCM has to take into account the chemical and geometrical 
structure of chain units that the system possesses \cite{georges,coulon1}. With 
quantum units, exact numerical diagonalization technique is not feasible, due to 
large size of the system, particularly so at low temperatures as correlation length 
grows exponentially as the temperature is lowered and the system size to be studied 
should at least have dimension of the correlation length. While a large number of 
SCMs have been synthesized over the years, here we will study an important class of 
SCMs in detail. In this class of SCMs, units are alternately isotropic and 
anisotropic. Furthermore, anisotropy axes are not collinear and since anisotropy is 
large and negative, the anisotropic units can be thought of as canted Ising spins 
at low temperatures. It is worth noting that, in this situation had the anisotropy 
axes been collinear, we could have used a simple one dimensional Ising model as only 
one component of the spins of the isotropic sites would be coupled to the anisotropic 
units. But the class of SCMs we wish to model, has non-collinearity of the anisotropy 
axes, therefore different components of the spin of the isotropic sites would be 
involved at different sites; as a result, we have to treat the isotropic spins 
quantum mechanically. 
Here we note that though all the interactions are quantum 
mechanical in nature, some of them become insignificant at low temperatures due to 
highly preferred spin-direction of some sites.
Herein, we model these type of systems exactly by transfer matrix method and study their 
low temperature statistical properties. We 
also discuss the issue of interchain interactions explicitly, which has been a long 
pending issue in this field \cite{miyasaka}.

There are some related theoretical/numerical studies which deserve a brief mention here. 
Alternative classical-quantum (or anisotropic-isotropic) spin systems have been studied 
previously \cite{dupas,seiden,verdaguer,georges1,curely,pelka,strecka}. There, 
classical nature of the spins is assumed from correspondence principle, due to large 
value of the spin S, of the quantum spins. In an another work, a helical spin chain 
with alternately classical spin (Ising-type) and quantum spin (coming from organic 
radical) spins is studied \cite{ravi}. But here the quantum spins are reduced to 
classical spins by approximating Ising direction by averaging directions of their 
neighboring classical spins.

Before we present our method in Sec. 3, we present in the next section (Sec. 2) a 
brief discussion of a real system which can be modeled to be in the class of SCMs we 
are interested here. This real example should clarify the present modeling issues and 
provide the motivation for the model studied in this paper. In Sec. 4, we apply our 
technique to the real system and present our results. We conclude our work in Sec. 5.

\section{Description of a Real System}
\begin{figure}[h]
\begin{center}
\hspace*{1cm}{\includegraphics[width=6.0cm]{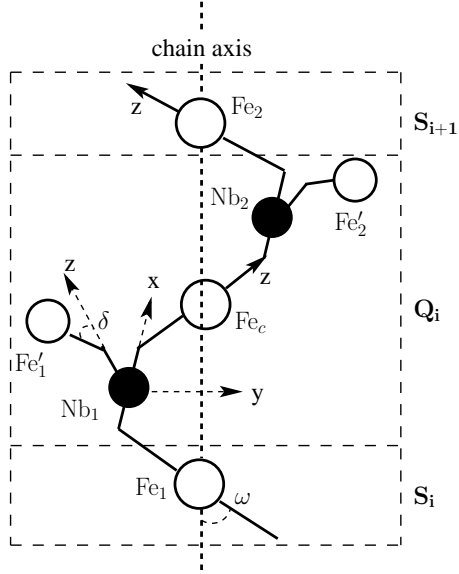}}
\caption{{\small One structural unit of $Fe_2Nb$ chain is shown here. $\omega$ is the 
angle made by anisotropy axis of in-chain $Fe$-ions with chain-axis. $\delta$ is the 
angle made by anisotropy axis of off-chain $Fe$ ion with $z$-axis of associated 
$Nb$-ion. $\alpha$ is the angle made by local $xy$-plane of Nb with the plane  
formed by all alternate in-chain Fe-ions (here $Fe_1$ and $Fe_2$) along with their 
anisotropy axes. Alternate classical and quantum units are also indicated here by 
dotted boxes (see Sec. 3).}}
\label{f1}
\end{center}
\end{figure}
Recently a compound, we refer to as $Fe_2Nb$, whose chemical formula is 
[\{$(H_2O)$ $Fe(L)$\} \{$Nb(CN)_8$\} \{$Fe(L)$\}]$_{\infty}$ has been synthesized 
\cite{venkat}. Its crystal structure studies reveal that it consists of topologically 
quasi-one dimensional spin chains with helical geometry. This system shows SCM 
character of slow relaxation and 
large DC susceptibility at low temperatures. The back bone of each chain consists of 
alternate $Fe$ and $Nb$ ions with each $Nb$ ion connected to an additional off-chain 
$Fe$ ion (see Fig \ref{f1}). $Nb$ ions have spin 1/2 while $Fe$ ions have spin of 2. 
The coordination around $Fe(II)$ is the unusual hepta coordination which leads to 
large anisotropy of the $Fe$ spins. All the intrachain interactions are 
antiferromagnetic and much stronger than interchain interaction. Since the $Fe$ ions 
have large and negative (easy-axis) anisotropy, we can assume the $Fe$ spins to be 
Ising-type, with non-collinear spin orientations due to helicity of the chain. As 
the spin axes are non-collinear, different components of $Nb$ spins are involved 
in interactions at different sites along the helix and 
therefore must be treated quantum mechanically. A careful examination of the chain 
structure, shows that two chemical units form a geometrical unit which are used as 
basic blocks in our transfer matrix method. To define this complex structural unit, 
we introduce three structural parameters, $\omega$, $\alpha$ and $\delta$, 
which are defined in Fig (\ref{f1}). We note that, here a quantum unit is formed by 
two $Nb$ ions connected by one in-chain $Fe$ ion and each of the $Nb$ ions being 
connected to an off-chain $Fe$ ion. These quantum units are connected by in-chain 
$Fe$ ions, which behave like classical spins. Therefore, this system falls into the 
class of SCMs we discussed in the previous section. In Sec 4, we first discuss in 
detail the application of our technique to solve the spin model for the SCM system. 
This is followed by presentation of results and their comparison with experimental 
studies.  

\section{Description of the Technique}
\begin{figure}[]
\begin{center}
\hspace*{1cm}{\includegraphics[width=12.0cm]{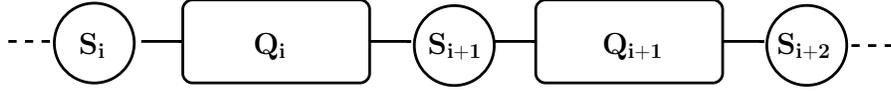}}
\caption{{\small Schematic diagram of a chain with alternative quantum and classical 
units}}
\label{f2}
\end{center}
\end{figure}

Let us consider our system formed by identical quantum units connected by classical 
(Ising-type) spins (see Fig \ref{f2}). Let ${\bf H}^{QM}_i({\bf Q}_i)$ be the 
Hamiltonian associated with $i$-th quantum unit with the set of quantum and classical 
spin operators ${\bf Q}_i$. Let the Hamiltonian for the interaction of the $i$-th 
quantum unit with classical spin, ${\bf S}_i$, to the left, be 
${\bf H}^{I,L}_i ({\bf q}^L_i, {\bf S}_i)$ with ${\bf q}^{L}_i$ the quantum spin 
operator of the $i$-th quantum unit involved in interaction to the left. We can 
similarly define ${\bf H}^{I,R}_i ({\bf q}^R_i, {\bf S}_{i+1})$ as the Hamiltonian 
for the interaction 
of the $i$-th quantum unit with classical spin, $S_{i+1}$, to the right. In the 
presence of (uniform) external magnetic field $\vec{B}$, there will be Zeeman 
interaction term $- g \mu_B {\bf S}_{Z,i} B$, where ${\bf S}_{Z,i}$ be the $Z$-component 
of spin for the 
$i$-th block, and this term is here interpreted as sum of the projection of spin for 
the $i$-th quantum unit and half of that for both $i$-th and $(i+1)$-th classical spins 
along the global (laboratory) $Z$-axis. Here ${\bf S}_{Z,i}$ is defined with weight 
1/2 for the classical spins to avoid double counting. 
We now can write our total Hamiltonian of a chain as, 
${\bf H} = \sum_{i=1}^N{\bf H}_i$, where, 
\begin{eqnarray}
{\bf H}_i & = &{\bf H}^{I,L}_i({\bf q}^L_i,{\bf S}_i) + {\bf H}^{QM}_i({\bf Q}_i) +
{\bf H}^{I,R}_i({\bf q}^R_i,{\bf S}_{i+1}) - g \mu_B {\bf S}_{Z,i} B,
\label{e2}
\end{eqnarray}
In the last term, the $g$-factor is assumed to be an average over all the corresponding 
$g$-factors of different spin-sites and $\mu_B$ is the Bohr magneton. We note that, the 
Zeeman interaction term has three different parts: 
$\frac{1}{2}{\bf H}^{Zm}_i({\bf S}_i,\vec{B})$, ${\bf H}^{Zm}_i({\bf Q}_i,\vec{B})$ and 
$\frac{1}{2}{\bf H}^{Zm}_{i+1}({\bf S}_{i+1},\vec{B})$ corresponding to the one associated 
with  $i$-th classical spin, $i$-th quantum unit and $(i+1)$-th classical spin respectively.
To avoid double counting, as before, Zeeman interaction part for $i$-th and $(i+1)$-th 
classical spins have been taken with weights 1/2.
Partition function for the chain (having $N$ quantum and classical units) at 
temperature $T$ can be written as,
\begin{eqnarray}
Q_N(\beta,B)={~\rm Tr ~} e^{-\beta {\bf H}}
\label{e3}
\end{eqnarray}
where, `Tr' means trace and $\beta = 1/(k_B T)$, $k_B$ being the Boltzmann constant. 
Now let us consider direct product basis of the whole chain. This can simply be 
written as, $|\cdots,\sigma_i,Q_i,\cdots \rangle$, where $|\sigma_i \rangle$ and 
$|Q_i \rangle$ represents convenient basis for $i$-th classical spin and quantum unit, 
respectively. Eq (\ref{e3}) can now be written as,
\begin{eqnarray}
Q_N(\beta,B)=\sum_{\{\sigma,Q\}} \langle \cdots,\sigma_i,Q_i,\cdots|e^{-\beta {\bf H}} 
|\cdots,\sigma_i,Q_i,\cdots \rangle
\label{e4}
\end{eqnarray}
Here the sums run over all possible configurations $\{\sigma,Q\}$ of all the classical 
spins and the quantum units. Now since, ${\bf H}_i$ and ${\bf H}_j$ commute, we can 
rewrite Eq (\ref{e4}) in the following form:
\begin{eqnarray}
Q_N(\beta,B)& = &\sum_{\{\sigma,Q\}} \langle \cdots,\sigma_i,Q_i,\cdots|\prod_{i=1}^N 
e^{-\beta {\bf H}_i} |\cdots,\sigma_i,Q_i,\cdots \rangle \nonumber \\
 & = & \sum_{\{\sigma\}}\langle \cdots,\sigma_i,\cdots|\prod_{i=1}^N~\left(\sum_{Q_i}
\langle Q_i|e^{-\beta {\bf H}_i}|Q_i\rangle\right)~|\cdots,\sigma_i,\cdots \rangle  
\label{e5}
\end{eqnarray}
Here, $Q_i$ is the set of configurations for $i$-th quantum unit. Note, the quantity 
in parentheses, $\displaystyle \sum_{Q_i}\langle Q_i|e^{-\beta {\bf H}_i}|Q_i\rangle$, 
is a purely classical operator (containing ${\bf S}_i$ and ${\bf S}_{i+1}$ only) as 
we have summed over quantum variables of the unit. Denoting it as ${\bf T}_i$ enables 
us to rewrite Eq (\ref{e5}) as,
\begin{eqnarray}
Q_N(\beta,B)& = & \sum_{\{\sigma\}}\langle \cdots,\sigma_i,\cdots|\prod_{i=1}^N~
{\bf T}_i~|\cdots,\sigma_i,\cdots \rangle
\label{e6}
\end{eqnarray}
This form is familiar to us, with ${\bf T}_i$ being the transfer operator. 
Introducing the identity $\displaystyle \sum_{\sigma_i} |\sigma_i \rangle \langle 
\sigma_i|$ between successive ${\bf T}$s, will enable us to write the Partition 
function as a trace of $N$-th power of a transfer matrix, ${\bf P}$, 
\begin{eqnarray}
Q_N(\beta,B)&=&{~\rm tr~}{\bf P}^N
\label{e7}
\end{eqnarray}
where, elements of the transfer matrix ${\bf P}$ are given by,
\begin{eqnarray}
P_{\sigma_i,\sigma_{i+1}}&=&\langle \sigma_i|{\bf T}_i|\sigma_{i+1}\rangle
\label{e8}
\end{eqnarray}
Since we have taken classical spins to be 
Ising-type (which can have only two values, $\pm 1$), the ${\bf P}$ matrix will be a  
$2\times2$ matrix. Diagonalizing this matrix is trivial; let $\lambda_+$ and 
$\lambda_-$ be the two eigenvalues of the matrix, then they can be expressed in terms 
of four elements of ${\bf P}$ as: $\lambda_{\pm}=\frac{1}{2}\left[(p_{11}+p_{22})\pm
\sqrt{(p_{11}-p_{22})^2+4p_{12}p_{21}}\right]$. In terms of eigenvalues, the partition 
function can now be written as,
\begin{eqnarray}
Q_N(\beta,B)&=&\lambda^N_+ + \lambda^N_-
\label{e9}
\end{eqnarray}
Note both the eigenvalues are function of $\beta$ and $B$. For thermodynamically 
large $N$, one can only take the larger eigenvalue (here $\lambda_+$), but for finite 
$N$, one should take both of them to evaluate the partition function. $\chi T$ as 
a function of $T$ ($\chi$ being susceptibility), can be obtained as, 
\begin{eqnarray}
\chi T&=&\frac{N_Ak_BT^2}{N}\frac{\partial^2}{\partial B^2} {~\rm ln~} Q_N(\beta,B)
\label{e10}
\end{eqnarray}
where, $N_A$ is the  Avogadro's number. In Eq (\ref{e10}), 2nd order differentiation 
can be done easily, one can even get simple closed analytical form for small applied 
magnetic field $B$. In general, one can do simple numerical differentiation to get 
$\chi T$ value as a function of temperature T. 

We now need to discuss a little more about ${\bf T}_i$. If we take basis 
$|Q_i \rangle$ 
to be eigenstates of ${\bf H}_i$, then ${\bf T}_i$ will just be sum of exponential of 
those eigenvalues. Let by diagonalizing it (if quantum unit has unconnected spin-1/2 
quantum spins, we can do it trivially and get analytical expression) we get 
eigenvalues $L^i_j$, where $j$ runs from $1$ to $d$ (number of basis vectors or 
dimensionality of configurational space for the quantum unit). Note these eigenvalues 
are functions of only ${\bf S}_i$ and ${\bf S}_{i+1}$, the classical spin operators. 
So we can write ${\bf T}_i$ in the following way:
\begin{eqnarray}
{\bf T}_i &=& \sum_{Q_i} \langle Q_i|e^{-\beta {\bf H}_i}|Q_i\rangle \nonumber \\
 &=&\sum_{j=1}^d e^{-\beta L^i_j} 
\label{e11}
\end{eqnarray}
In our case ($Fe_2Nb$), quantum unit has five spins (3 Ising spins and 2 quantum 
spin-1/2s), hence $d$ = 32 and ${\bf T}_i$ will involve summation over the 32 
eigenvalues.

\subsection{Consideration of Interchain Interaction} 
Interchain interaction is an important issue in this field of SCMs; in real systems 
it is not possible to separate chains to eliminate this interaction. With increase in 
this interaction strength, a one dimensional system like SCM gradually transforms 
into a higher dimensional system with remarkably different properties. Even a 
weak interchain interaction can change properties of a chain -its presence can be 
seen both in static (like $\chi T vs. T$ plots, hysteresis curves) and dynamic (like, 
Cole-Cole plots) measurements. Unfortunately, there is no detailed and explicit 
discussion on how one can treat this interaction theoretically. In fact, this issue 
has been highlighted in a paper by Miyasaka {\it et al.} \cite{miyasaka}, where it is 
posed as an open problem. In this section we discuss interchain interaction in detail. 

In low-dimension, the quantum fluctuations are generally large. However SCMs are 
not strictly one-dimensional, and we have weak interchain interaction. It is 
customary to deal with interchain interactions within a mean field approximation.
Here a widely used formula for corrected (or modified) susceptibility due to the 
interaction is $\chi_M=\chi_0/(1+zJ'\chi_0/N_Ag^2\mu^2_B)$, where $\chi_0$ is 
susceptibility of the chain without interchain interactions, $z$ is the number of 
nearest neighbors and $J'$ is interchain interaction strength ($zJ'$ together is 
called mean field parameter) \cite{ferbinteanu,clerac,zheng}. However we wish to 
emphasize that, this formula is for vanishing applied magnetic field. So in case of 
finite magnetic field and at very low temperatures this formula is not appropriate. 
In addition, for a chain, unlike for molecular systems, $S_Z$ can be very large; so 
it is not clear whether it is valid to make series expansion of exponential function 
of the Zeeman and interchain interactions and retain terms only up to first order. 
Here we discuss a mean-field approach where the interaction is dealt by 
transfer matrix technique and can be used to study the case when magnetic field is 
finite. We also wish to make it clear that, there has to be a factor $N$ (number 
of chemical units in a chain) multiplying $zJ'$ in the above formula of modified 
susceptibility for vanishing magnetic field. To the best of our knowledge, this is 
the first attempt to deal with interchain interactions by the transfer matrix 
technique.

If the interchain interaction is weak, we can bring in the effect of environment by 
placing a chain  in the mean field generated by neighboring chains. Let us now 
consider $\langle S_Z \rangle$ be the expectation value of $Z$-component of spin for 
each chain and $zJ'$ be the interchain interaction (mean field) parameter, then the 
interaction energy can be written as: $zJ' {\bf S}_Z{\langle S_Z \rangle}$, with 
${\bf S}_Z$ being the $Z$-component of total spin for a chain. Now since 
${\displaystyle {\bf S}_Z=\sum^N_{i=1} {\bf S}_{Z,i}}$ with ${\bf S}_{Z,i}$ being the 
$Z$-component of spin for $i$-th block, we can rewrite the interaction energy 
as  ${\displaystyle \sum^N_{i=1} (zJ' {\bf S}_{Z,i}{\langle S_Z \rangle})}$, which 
can be viewed as sum of interaction energy of each block of a chain with a mean field 
generated by neighboring chains. Now if we add this mean field block interaction 
energy term, $zJ' {\bf S}_{Z,i}{\langle S_Z \rangle}$, to ${\bf H}_i$ of 
Eq (\ref{e2}), we get,  
\begin{eqnarray}
{\bf H}_i & = &{\bf H}^{I,L}_i({\bf q}^L_i,{\bf S}_i) + {\bf H}^{QM}_i({\bf Q}_i) +
{\bf H}^{I,R}_i({\bf q}^R_i,{\bf S}_{i+1}) - g \mu_B {\bf S}_{Z,i} B + 
zJ' {\bf S}_{Z,i}{\langle S_Z \rangle} \nonumber\\
 & = & {\bf H}^{I,L}_i({\bf q}^L_i,{\bf S}_i) + {\bf H}^{QM}_i({\bf Q}_i) +
{\bf H}^{I,R}_i({\bf q}^R_i,{\bf S}_{i+1}) - g \mu_B {\bf S}_{Z,i} B' 
\label{e12}
\end{eqnarray}
where, $B' = B - \frac{zJ'}{g \mu_B}{\langle S_Z \rangle}$. We can then follow the 
usual transfer matrix technique and get a partition function for the chain in the 
presence of a modified magnetic field $B'$. Since, 
$\langle S_Z \rangle = \frac{1}{g\mu_B \beta} \frac{\partial {\rm ln~} Q_N(\beta,B')} 
{\partial B'}$, we see that, $\langle S_Z \rangle$ has to be solved self-consistently 
by an iterative procedure. We can use the following self-consistent scheme 
developed by first noting that, $zJ'$ being very small, internal field, $\Delta B = - 
\frac{zJ'} {g \mu_B} {\langle S_Z \rangle}$, is a slowly varying function of external 
applied field $B$ resulting in $|\frac{\partial \Delta B}{\partial B}|\ll 1$ or
$|\frac{\partial B'}{\partial B}|\approx 1$. Using the conditions and making 
the Taylor series expansion of ${\rm ln~} Q_N(\beta,B')$ about $B$, we get, 
\begin{eqnarray}
g\mu_B \beta~\langle S_Z \rangle &=& \frac{\partial{\rm ln~}Q_N(\beta,B')}
{\partial B'}\nonumber \\
 &\simeq&\frac{\partial{\rm ln~}Q_N(\beta,B)}{\partial B} - \left(\frac{zJ'}
{g\mu_B}\right)\frac{\partial^2{\rm ln~}Q_N(\beta,B)}{\partial B^2}
\langle S_Z \rangle + \nonumber \\ 
& &~\frac{1}{2}\left(\frac{zJ'}{g\mu_B}\right)^2
\frac{\partial^3{\rm ln~}Q_N(\beta,B)}{\partial B^3}\langle S_Z \rangle ^2 - \cdots
\label{e13}
\end{eqnarray}
Iteration can be started with $\langle S_Z \rangle = \frac{1}{g\mu_B \beta}
\frac{\partial {\rm ln~} Q_N(\beta,B)}{\partial B}$, without interchain 
interaction, finally leading to a converged $\langle S_Z \rangle$ value. The 
self-consistent result correct to second order in $zJ'$ is given by, 
\begin{eqnarray}
g\mu_B\langle S_Z \rangle &=& M(\beta,B) -\left(\frac{zJ'}{g^2\mu^2_B}\right)
M(\beta,B)\frac{\partial M(\beta,B)}{\partial B}+ \nonumber\\
 & &~\left(\frac{zJ'}{g^2\mu^2_B}\right)^2\left[M(\beta,B) \left(\frac{\partial 
M(\beta,B)}{\partial B}\right)^2+\frac{1}{2}(M(\beta,B))^2\frac{\partial^2 M(\beta,B)}
{\partial B^2}\right]
\label{e14}
\end{eqnarray}
where, $M(\beta,B)=\frac{1}{\beta}\frac{\partial {\rm ln~} Q_N(\beta,B)}
{\partial B}$ is the magnetization in the absence of interchain interaction. 
Going to higher order in $zJ'$ is possible but generally not necessary. Obtaining 
modified susceptibility ($\chi_M$) from Eq (\ref{e14}) is straight forward. In the 
vanishing field regime, we can get the widely used formula for $\chi_M$ by retaining 
terms up to first order in $zJ'$ of the Eq (\ref{e13}) and subsequently solving for 
$\langle S_Z \rangle$. This gives,
\begin{eqnarray}
g\mu_B\langle S_Z \rangle = \frac{\frac{1}{\beta}\frac{\partial{\rm ln~}Q_N(\beta,B)}
{\partial B}}{1+\frac{zJ'}{g^2\mu^2_B\beta}\frac{\partial^2{\rm ln~}Q_N(\beta,B)}
{\partial B^2}}
\label{e15}
\end{eqnarray}
Now in this regime, modified susceptibility is $\chi_M= \frac{N_A} {BN}g\mu_B\langle 
S_Z \rangle$ and susceptibility without interchain interaction is 
$\chi_0 = \frac{N_A}{B N\beta}\frac{\partial{\rm ln~}Q_N(\beta,B)}{\partial B}$. 
This directly leads to:
\begin{eqnarray}
\chi_M=\frac{\chi_0}{1+\frac{zJ'N}{g^2\mu^2_BN_A}\chi_0}
\label{e16}
\end{eqnarray}
Note the difference of factor of $N$ with this formula and the one generally used 
in this field \cite{ferbinteanu,clerac,zheng}. Appearance of this factor in 
Eq (\ref{e16}) is easy to understand: while susceptibility is measured for Avogadro 
number ($N_A$) of chemical units, in the transfer matrix method we compute partition 
function for $N$ (chemical) units; therefore, we need to normalize the susceptibility 
computed from the partition function. The factor $NzJ'$ can be given a physical 
interpretation. Let us assume that a unit in a given chain interacts only with the 
corresponding units of the $z$ neighboring chains and $p$ be the total effective 
coupling strength, then the associated interaction term will be $p {\bf S}_{Z,i} 
\frac{\langle S_Z \rangle}{N}$, where $\frac{\langle S_Z \rangle}{N}$ is the 
expectation value of $Z$-component of spin for a unit. The total mean field 
interaction term is therefore given by $\sum_i p {\bf S}_{Z,i} 
\frac{\langle S_Z \rangle}{N}$. Comparing this with conventional chain-chain total 
interaction term $\sum_i zJ' {\bf S}_{Z,i}\langle S_Z \rangle$, we get $p = NzJ'$. 
Thus, while $zJ'$ is the coupling strength of a chain with its $z$ neighboring 
chains, $NzJ'$ is the coupling strength of a given unit with the corresponding units 
of the $z$ neighboring chains. We will report the latter parameter ($NzJ'$) for our 
study of real system.

This way of introducing interchain interaction in computing magnetic susceptibility 
is fairly general and can be applied to any class of SCMs. Here we have presented a 
formula for modified susceptibility which is very similar to the one used for 
molecular systems and generally adopted for SCMs. The higher order self-consistent 
method (Eq \ref{e14}) is more general and applicable even when the magnetic field is 
not vanishingly small. 

\section{Application to the $Fe_2Nb$ system}
In this section we will apply our technique to the system $Fe_2Nb$ which is already 
introduced in Sec. 2. All the $Fe-Nb$ couplings are supposed to be antiferromagnetic 
and are assumed to be of same strength ($J$). We assume that the strength 
of axial anisotropy constant $D$ to be the same for each $Fe$ ions. As we already 
mentioned, since this $D$ is known to be large and negative, we treat the spins 
on $Fe$ ions as two state Ising spins along their respective anisotropy axis. We also 
further assume the $g$-factor to be isotropic and has same value for all the ions. 
Three parameters are defined in Fig (\ref{f1}), namely, $\omega$, 
$\alpha$ and $\delta$, to describe complex structural unit of the chain. In 
Eq (\ref{e10}) we need to use average partition function, averaged over all possible 
orientations of a chain to compare with experiments done on a powder sample. The 
averaging can be done by noting that a chain orientation in space can be defined by 
three angles, namely, $\theta$, $\phi$ and $\psi$. Here $\theta$ and $\phi$ are usual 
spherical polar coordinates to define the orientation of the chain axis and $\psi$ is 
measure of rotation of the chain about its own axis. Our partition function will 
depend on (a) temperature, (b) magnetic field and (c) all three structural parameters 
($\omega$, $\alpha$ and $\delta$)for a single chain, it will also depend on 
orientational angles ($\theta$, $\phi$ and $\psi$) with respect to laboratory frame. 
The partition function can be averaged over all orientations of the chain axis to 
obtain an average partition function $\bar{Q}_N(\beta,B)$, given by,
\begin{eqnarray}
\bar{Q}_N(\beta,B)=\int_{\theta=0}^{\pi}\int_{\phi=0}^{2\pi}\int_{\psi=0}^{2\pi}
Q_N(\beta,B;\theta,\phi,\psi){~\rm sin}\theta~d\theta~d\phi~d\psi
\label{e17}
\end{eqnarray}
where, $Q_N(\beta,B;\theta,\phi,\psi)$ is the partition function of the chain for a 
particular orientation ($\theta,\phi,\psi$). We have ignored a normalization 
constant, as it will not affect our result. Integrations in Eq (\ref{e17}) can be 
done by simple numerical methods (analytically it may not always be possible). So in 
this particular problem, Eq (\ref{e10}) must be replaced by, 
\begin{eqnarray}
\chi T&=&\frac{N_Ak_BT^2}{2N}\frac{\partial^2}{\partial B^2} {~\rm ln~} 
{\bar Q}_N(\beta,B)
\label{e18}
\end{eqnarray}
Note, we have also replaced $N$ by $2N$, as one structural unit consists of two 
chemical units and susceptibility from experiment is quoted for an Avogadro number of 
chemical units.
 
The Hamiltonian for $i$-th quantum unit consisting of $Nb_1,~Fe'_1,~Fe_c,~Nb_2$ and 
$Fe'_2$ site spins and connected to neighboring quantum units by $Fe_1$ and 
$Fe_2$ site spins respectively, can be written as: 
\begin{eqnarray}
{\bf H}_i &=& J\vec{\bf{S}}^{Fe_1}\cdot\vec{\bf{S}}^{Nb_1} + 
J\vec{\bf{S}}^{Fe_2}\cdot\vec{\bf{S}}^{Nb_2} +
J(\vec{\bf{S}}^{Fe'_1}\cdot\vec{\bf{S}}^{Nb_1} + 
\vec{\bf{S}}^{Fe_c}\cdot\vec{\bf{S}}^{Nb_1} +
\vec{\bf{S}}^{Fe_c}\cdot\vec{\bf{S}}^{Nb_2} +
\vec{\bf{S}}^{Fe'_2}\cdot\vec{\bf{S}}^{Nb_2})\nonumber \\
 & & -\frac{1}{2}g\mu_B\vec{\bf{S}}^{Fe_1}\cdot\vec{B}- 
\frac{1}{2}g\mu_B\vec{\bf{S}}^{Fe_2}\cdot\vec{B} \nonumber \\
 & & -g\mu_B(\vec{\bf{S}}^{Nb_1}\cdot\vec{B} + \vec{\bf{S}}^{Fe'_1}\cdot\vec{B} +
\vec{\bf{S}}^{Fe_c}\cdot\vec{B} + \vec{\bf{S}}^{Nb_2}\cdot\vec{B} +
\vec{\bf{S}}^{Fe'_2}\cdot\vec{B})
\label{e19}
\end{eqnarray}
Here $\vec{\bf{S}}$ is a vector spin operator for the specified ion. All these spin 
operators are defined relative to their corresponding local coordinate systems (see 
Fig \ref{f1}). For example, $z$-axis of a $Fe$ ion is taken along its easy anisotropy 
axis. First and second terms in Eq (\ref{e19}) are interaction terms between quantum 
unit and its neighboring classical units. The 3rd term within brackets includes all 
the interactions within the quantum unit. The 4th, 5th and 6th terms are the Zeeman 
terms associated with classical spins with appropriate weight and the quantum unit 
respectively. All these terms can be expressed explicitly with the help of structural 
angles or parameters ($\omega,~\alpha,~\delta$) and polar coordinates of the chain 
($\theta,~\phi,~\psi$). Table (\ref{tbl1}) gives the expression for each of the terms 
in the laboratory frame assuming that the magnetic field is applied along $Z$-axis 
of the laboratory frame.
\begin{table}[htbp]
\caption{{\small Expression for all 13 interaction terms appearing in Eq (\ref{e19})} 
is given below. Treating the $Fe$ spins as Ising spins with anisotropy along $z$
direction, we replace the spin by $2\sigma^{Fe_1}$, where $\sigma^{Fe_1}$ is the 
Ising-variable with only possible values $\pm 1$, the factor 2 is to account for the 
spin of the $Fe$ ion which is 2.}
\begin{center}
\begin{tabular}{|l|l|}
\hline
Term & Expression \\ \hline
$\vec{\bf{S}}^{Fe_1}\cdot\vec{\bf{S}}^{Nb_1}$ & $2\sigma^{Fe_1}({\bf S}^{Nb_1}_x
{~\rm cos}\omega{~\rm cos}\alpha + {\bf S}^{Nb_1}_y{~\rm sin}\omega +
{\bf S}^{Nb_1}_z{~\rm cos}\omega{~\rm sin}\alpha)$ \\ \hline

$\vec{\bf{S}}^{Fe_2}\cdot\vec{\bf{S}}^{Nb_2}$ & $2\sigma^{Fe_2}({\bf S}^{Nb_2}_x
{~\rm cos}\omega{~\rm cos}\alpha - {\bf S}^{Nb_2}_y{~\rm sin}\omega +
{\bf S}^{Nb_2}_z{~\rm cos}\omega{~\rm sin}\alpha)$ \\ \hline

$\vec{\bf{S}}^{Fe'_1}\cdot\vec{\bf{S}}^{Nb_1}$ & $2\sigma^{Fe'_1}({\bf S}^{Nb_1}_z
{~\rm cos}\delta + {\bf S}^{Nb_1}_y {~\rm sin}\delta)$ \\ \hline

$\vec{\bf{S}}^{Fe_c}\cdot\vec{\bf{S}}^{Nb_1}$ & $2\sigma^{Fe_c}({\bf S}^{Nb_1}_x 
{~\rm cos}\omega{~\rm cos}\alpha -  {\bf S}^{Nb_1}_y{~\rm sin}\omega +
{\bf S}^{Nb_1}_z{~\rm cos}\omega{~\rm sin}\alpha)$ \\ \hline

$\vec{\bf{S}}^{Fe_c}\cdot\vec{\bf{S}}^{Nb_2}$ & $2\sigma^{Fe_c}({\bf S}^{Nb_2}_x
{~\rm cos}\omega{~\rm cos}\alpha + {\bf S}^{Nb_2}_y{~\rm sin}\omega +
{\bf S}^{Nb_2}_z{~\rm cos}\omega{~\rm sin}\alpha)$ \\ \hline

$\vec{\bf{S}}^{Fe'_2}\cdot\vec{\bf{S}}^{Nb_2}$ & $2\sigma^{Fe'_2}({\bf S}^{Nb_2}_z
{~\rm cos}\delta + {\bf S}^{Nb_2}_y {~\rm sin}\delta)$ \\ \hline

$\vec{\bf{S}}^{Fe_1}\cdot\vec{B}$ & $2\sigma^{Fe_1}({~\rm cos}\omega~{~\rm cos}\theta 
+ {~\rm sin}\omega~{~\rm sin}\psi~{~\rm sin}\theta)B$ \\ \hline

$\vec{\bf{S}}^{Fe_2}\cdot\vec{B}$ & $2\sigma^{Fe_2}({~\rm cos}\omega~{~\rm cos}\theta
+ {~\rm sin}\omega~{~\rm sin}\psi~{~\rm sin}\theta)B$ \\ \hline

$\vec{\bf{S}}^{Nb_1}\cdot\vec{B}$ & $[{\bf S}^{Nb_1}_x({\rm cos}\alpha~{\rm cos}\theta
-{\rm sin}\alpha~{\rm sin}\theta~{\rm cos}\psi) + {\bf S}^{Nb_1}_y{\rm sin}\psi
~{\rm sin}\theta$  \\
 & $+{\bf S}^{Nb_1}_z({\rm sin}\alpha~{\rm cos}\theta +
{\rm cos}\alpha~{\rm sin}\theta~{\rm cos}\psi)]B$ \\ \hline

$\vec{\bf{S}}^{Fe'_1}\cdot\vec{B}$ & $2\sigma^{Fe'_1}({\rm sin}\delta~{\rm sin}\psi~
{\rm sin}\theta + {\rm cos}\delta~{\rm sin}\alpha~{\rm cos}\theta + {\rm cos}\delta~ 
{\rm cos}\alpha~{\rm cos}\psi~{\rm sin}\theta)B$ \\ \hline

$\vec{\bf{S}}^{Fe_c}\cdot\vec{B}$ & $2\sigma^{Fe_c}({\rm cos}\omega~{\rm cos}\theta - 
{\rm sin}\omega~{\rm sin}\psi~{\rm sin}\theta)B$ \\ \hline

$\vec{\bf{S}}^{Nb_2}\cdot\vec{B}$ & $[{\bf S}^{Nb_2}_x({\rm cos}\alpha~{\rm cos}\theta
+{\rm sin}\alpha~{\rm sin}\theta~{\rm cos}\psi) - {\bf S}^{Nb_2}_y{\rm sin}\psi
~{\rm sin}\theta$  \\
 & $+{\bf S}^{Nb_2}_z({\rm sin}\alpha~{\rm cos}\theta -
{\rm cos}\alpha~{\rm sin}\theta~{\rm cos}\psi)]B$ \\ \hline

$\vec{\bf{S}}^{Fe'_2}\cdot\vec{B}$ & $2\sigma^{Fe'_2}(-{\rm sin}\delta~{\rm sin}\psi~
{\rm sin}\theta + {\rm cos}\delta~{\rm sin}\alpha~{\rm cos}\theta - {\rm cos}\delta~
{\rm cos}\alpha~{\rm cos}\psi~{\rm sin}\theta)B$ \\ \hline

\end{tabular}
\end{center}
\label{tbl1}
\end{table}
Knowing all these interaction terms, now we can form the transfer operator ${\bf T}$ 
for the $i$-th unit as,
\begin{eqnarray}
{\bf T}_i&=&\sum_{\{Q_i\}} \langle Q_i|e^{-\beta {\bf H}_i}|Q_i\rangle \nonumber \\ 
 &=& \sum_{\{Q_i\}} \langle Q_i|e^{(a{\bf S}^{Nb_1}_x+b{\bf S}^{Nb_1}_y+
c{\bf S}^{Nb_1}_z)}e^{(d{\bf S}^{Nb_2}_x+e{\bf S}^{Nb_2}_y+f{\bf S}^{Nb_2}_z)}
e^h|Q_i \rangle
\label{e22}
\end{eqnarray}
Expressions for $a$, $b$, $c$, $d$, $e$, $f$ and $h$, which are functions of 
classical spins involved in the problem, are given in Table (\ref{tbl2}). 
Here $|Q_i\rangle$ 
\begin{table}[htbp]
\caption{{\small Expression for variables appeared in Eq (\ref{e22})}}
\begin{center}
\begin{tabular}{|c|l|}
\hline
Variable & Expression \\ \hline
$a$ & $-\beta[2J(\sigma^{Fe_1}+\sigma^{Fe_c}){\rm cos}\omega~{\rm cos}\alpha-g\mu_BB
({\rm cos}\alpha~{\rm cos}\theta-{\rm sin}\alpha~{\rm sin}\theta~{\rm cos}\psi)]$ \\ 
\hline
$b$ & $-\beta[2J\{(\sigma^{Fe_1}-\sigma^{Fe_c}){\rm sin}\omega+
\sigma^{Fe'_1}{\rm sin}\delta\}-g\mu_BB~{\rm sin}\psi~{\rm sin}\theta]$\\ \hline
$c$ & $-\beta[2J\{(\sigma^{Fe_1}+\sigma^{Fe_c}){\rm cos}\omega~{\rm sin}\alpha +
\sigma^{Fe'_1}{\rm cos}\delta\} -g\mu_BB({\rm sin}\alpha~{\rm cos}\theta+
{\rm cos}\alpha~{\rm sin}\theta~{\rm cos}\psi)]$\\ \hline
$d$ & $-\beta[2J(\sigma^{Fe_c}+\sigma^{Fe_2}){\rm cos}\omega~{\rm cos}\alpha-g\mu_BB
({\rm cos}\alpha~{\rm cos}\theta+{\rm sin}\alpha~{\rm sin}\theta~{\rm cos}\psi)]$\\ 
\hline
$e$ & $-\beta[2J\{(\sigma^{Fe_c}-\sigma^{Fe_2}){\rm sin}\omega+
\sigma^{Fe'_2}{\rm sin}\delta\}+g\mu_BB~{\rm sin}\psi~{\rm sin}\theta]$\\\hline
$f$ & $-\beta[2J\{(\sigma^{Fe_c}+\sigma^{Fe_2}){\rm cos}\omega~{\rm sin}\alpha +
\sigma^{Fe'_2}{\rm cos}\delta\} -g\mu_BB({\rm sin}\alpha~{\rm cos}\theta-
{\rm cos}\alpha~{\rm sin}\theta~{\rm cos}\psi)]$\\ \hline
$h$ & $\beta g\mu_BB[(\sigma^{Fe_1}+\sigma^{Fe_2})({\rm cos}\omega~{\rm cos}\theta+
{\rm sin}\omega~{\rm sin}\theta~{\rm sin}\psi)$\\
 &$ +2\sigma^{Fe'_1}({\rm sin}\delta~{\rm sin}\theta~{\rm sin}\psi+{\rm cos}\delta~
{\rm sin}\alpha~{\rm cos}\theta+{\rm cos}\delta~{\rm cos}\alpha~{\rm sin}\theta~
{\rm cos}\psi)$\\
 &$ +2\sigma^{Fe_c}({\rm cos}\omega~{\rm cos}\theta-{\rm sin}\omega~{\rm sin}\theta~
{\rm sin}\psi)$\\
 &$ +2\sigma^{Fe'_2}(-{\rm sin}\delta~{\rm sin}\theta~{\rm sin}\psi+{\rm cos}\delta~
{\rm sin}\alpha~{\rm cos}\theta-{\rm cos}\delta~{\rm cos}\alpha~{\rm sin}\theta~
{\rm cos}\psi)]$ \\ \hline
\end{tabular}
\end{center}
\label{tbl2}
\end{table}
is the direct product basis of the spins of $Nb_1$, $Fe'_1$, $Fe_c$, 
$Nb_2$, and $Fe'_2$. If we take the basis for $Nb_1$ to be eigenstates of 
($a{\bf S}^{Nb_1}_x+b{\bf S}^{Nb_1}_y+c{\bf S}^{Nb_1}_z$) (the quantum mechanical 
operator in the exponent of the first exponential in Eq (\ref{e22})), the 
corresponding exponential term will be reduced to sum of exponential of the 
eigenvalues of the operator. Similarly second exponential term can also be reduced to 
the sum of exponential of the eigenvalues of the corresponding quantum operator 
($d{\bf S}^{Nb_2}_x+e{\bf S}^{Nb_2}_y+f{\bf S}^{Nb_2}_z$). Finding eigenvalues of 
these quantum mechanical operators is not difficult; they are $\pm \lambda^{Nb_1}$ 
and $\pm \lambda^{Nb_2}$ respectively, where, 
$\lambda^{Nb_1}=\frac{1}{2}\sqrt{a^2+b^2+c^2}$ and 
$\lambda^{Nb_2}=\frac{1}{2}\sqrt{d^2+e^2+f^2}$. We now write the transfer 
operator in Eq (\ref{e22}) as,
\begin{eqnarray}
{\bf T}_i&=&\sum_{\{Q'_i\}} (e^{\lambda^{Nb_1}}+e^{-\lambda^{Nb_1}})
(e^{\lambda^{Nb_2}}+e^{-\lambda^{Nb_2}})e^h \nonumber \\
& =&4\sum_{\{Q'_i\}}{\rm cosh}\lambda^{Nb_1}~{\rm cosh}\lambda^{Nb_2}~e^h
\label{e23}
\end{eqnarray}
where the sum runs over all the configurations of the classical spins, 
$\{\sigma^{Fe'_1},\sigma^{Fe_c},\sigma^{Fe'_2}\}$, within the quantum unit $i$, which 
is denoted by $\{Q'_i\}$. Note, ${\bf T}_i$ depends only on ${\sigma^{Fe_1}}$ and 
${\sigma^{Fe_2}}$ and we can write elements of transfer matrix {\bf P} as:
\begin{eqnarray}
p_{\sigma^{Fe_1},\sigma^{Fe_2}}&=&4\sum_{\{Q'_i\}}{\rm cosh}\lambda^{Nb_1}~{\rm cosh}
\lambda^{Nb_2}~e^h
\label{e24}
\end{eqnarray}
where $\lambda^{Nb_1}$, $\lambda^{Nb_2}$ and $h$ are evaluated with the particular 
value of ($\sigma^{Fe_1},\sigma^{Fe_2}$). Next step here would be to get eigenvalues 
of the {\bf P} matrix, which in turn will give partition function for a particular 
orientation ($\theta,\phi,\psi$) of the chain. Now Eq (\ref{e17}), Eq (\ref{e18}) and 
Eq (\ref{e16}) can be used for obtaining $\chi T vs. T$ curve at low temperature 
(note, while Eq (\ref{e18}) will give $\chi_0$, Eq (\ref{e16}) will give us 
$\chi_M$).   
 
\subsection{Result and Discussion}
\begin{figure}[t]
\begin{center}
\hspace*{1cm}{\includegraphics[width=12.0cm]{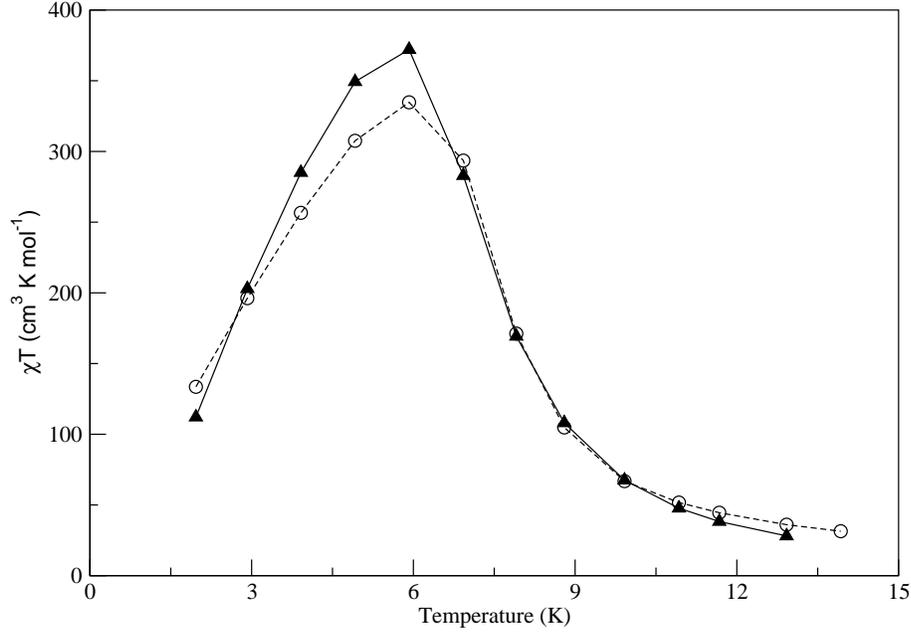}}
\caption{{\small $\chi T vs. T$ curve for $Fe_2Nb$. Open circles ($\ocircle$), 
connected by segmented line, represent experimental data points whereas filled 
triangles ($\blacktriangle$), connected by continuous line, represent calculated 
data points. The experiment was done with applied field of 100 Oe. All the 
theoretical points are calculated with same value of applied field.}}
\label{f3}
\end{center}
\end{figure}
In  Fig (\ref{f3}) we show the best fit $\chi T vs. T$ plot with experimental data. 
The best fit parameters are: $J$ = 20.3 $cm^{-1}$ (antiferromagnetic), $NzJ'$ = 
2.3$\times$10$^{-4}$ $cm^{-1}$ (antiferromagnetic), $g$ = 2.08, 
$\alpha$ = 33$^\circ$, $\omega$ = 22.5$^\circ$ and $\delta$ = 8$^\circ$. The 
theoretical fit was obtained from 110 geometrical units or 220 chemical units. We 
observe that, below 6K, $\chi T$ value decrease with temperature due to two effects: 
(a) finite size of chains and (b) antiferromagnetic interchain interaction. Due to 
finite size, magnetization of the chain has an upper bound and $\chi$ reaches a 
saturation value and below this temperature, $\chi T$ value will begin to decrease. 
Effect of antiferromagnetic interchain interaction further reduces the value of 
susceptibility at low temperatures. We also note that, at higher temperatures
theoretical value of susceptibility is lower compared to the experimental value. 
Since, unlike in a real system, the direction of $Fe$ moments in the model are fixed  
and can not change, resulting in smaller magnetization than the powder compound. 
Two more things can be estimated from our model; first is the energy for creating a 
domain wall in the chain, $\Delta_{\xi}$, the second being the anisotropy constant, 
$D$.

One of the important properties of SCMs is the slow relaxation of magnetization. The 
relaxation dynamics is controlled by the blocking temperature of the system and is 
related to the energy required to create a domain wall in the chain. The lowest 
energy required to create a domain wall is the absolute difference in energy between 
the ground state and an exited state, in which one part of the chain is magnetized in 
one direction and the rest of the chain is magnetized in the opposite direction. In 
the class of SCMs we are discussing, visualization of domain wall may be difficult 
due to its complex structure, especially due to the presence of both quantum and 
classical units. It is sufficient here to consider each $Nb$ ion and the associated 
off-chain $Fe$ ion as one quantum unit, since all of these units are identical in the 
absence of external magnetic field. In order to get ground state energy of the SCM, 
we solve for the eigenvalues of a quantum unit for different configurations of the 
surrounding classical units. Here the Hamiltonian to be considered includes 
interaction of 
$Nb$ ion with the associated off-chain $Fe$ ion and half of the interactions of that 
$Nb$ ion with surrounding in-chain $Fe$ ions. We obtain the minimum energy when all 
the three Ising spins connected to the $Nb$ ion are in same state, say,`+1' state. So 
the ground state of the chain is when all the Ising spins (both in- and off-chin) are 
in the `+1' state. Now to create a domain wall, we reverse all the Ising spins of one 
part, say right half, and keep the remaining Ising spins of left half in the initial 
state of `+1'. This will change energy of only one quantum unit, where the off-chain 
and one in-chain Ising spins are in `+1' state and another in-chain Ising spin is in 
`-1' state. The difference in energy, i.e., the energy to create a domain wall is 
given by (see Fig \ref{f1}):
\begin{eqnarray}
\Delta_{\xi}&=&\frac{1}{2}\left(\sqrt{a^2+b^2+c^2}~\big|_{\sigma^{Fe_c}=1}-
\sqrt{a^2+b^2+c^2}~\big|_{\sigma^{Fe_c}=-1}\right)_{\sigma^{Fe_1}=1,\sigma^{Fe'_1}=1}
\nonumber \\
&\approx& 34.5 {\rm K}
\label{e25}
\end{eqnarray}
This compares well with the experimental energy gap of 33 K obtained from 
log($\chi T$) vs. $1/T$ plot above temperature of 6K (to exclude effects due to 
finite size and interchain interactions). 

Estimation of anisotropy parameter ($D$) can not be done directly from the fits as 
our Hamiltonian does not contain the $D$ parameter. Here we employ an indirect way for 
its estimation. We have assumed that low temperature spins of $Fe$-ions behave like 
Ising spins due to large and negative $D$-value from experiment. Spin of $Fe$ ion 
being 2, we have also assumed that spin values that can be assumed by $Fe$ is only 
$\pm$2. If we attribute the deviation of the theoretical curve from the experimental 
curve due to an assumption of canted Ising-type spin for the $Fe$ spins, then from 
the deviation temperature $T_d$ (about 12K in Fig \ref{f3}) we can estimate the 
anisotropy parameter. To do this, we solve the quantum mechanical exchange 
Hamiltonian, namely, $H = J\vec{\bf S}^{Fe}\cdot\vec{\bf S}^{Nb} - 
D({\bf S}^{Fe}_Z)^2$, with both $J$ and $D$ positive, of 2-spin problem involving 
$Fe$ and $Nb$ spins exactly. Note, all the intrachain interactions are of this 
antiferromagnetic $Fe-Nb$ type. We conjecture that the deviation from the 
experimental $\chi T vs. T$ plot occurs due to significant population of the 3rd 
eigenstate relative to the second eigenstate. That is the thermal energy $k_BT_d$ is 
such that the ratio of the population of the 3rd eigenstate to the 2nd eigenstate is 
$1/e$. This yields the result that $\Delta E = E_3-E_2 \approx 3D - 2J = k_B T_d$. 
Since we know $J$ from the fitting, we estimate $D \sim$ 16.3 $cm^{-1}$. This 
estimation is close to the experimental value of 17 $cm^{-1}$.
 
\section{Conclusion}
We have outlined a method which can be used to study thermodynamic properties
of an important class of SCMs where 
alternate units are of classical and quantum nature with anisotropy axes 
of classical units being non-collinear. We have also discussed in detail the 
issue of interchain interaction and showed how to incorporate it in transfer matrix 
technique. We applied this method to a real system and studied its low 
temperature behavior. We have also carried out an averaging process that allows computing 
the magnetic properties of a powder sample. We have estimated anisotropy parameter 
and energy associated with domain wall within this model. At this point it is worth 
mentioning that, it may be better to consider all five states (0, $\pm$1, $\pm$2) of 
the $z$-component of $Fe$ spin instead of the two states considered here. However, in 
that case, one has to deal with little bigger transfer matrix (dimension 5$\times$5); 
but the main disadvantage of this extension will be that it will add one more 
parameter to the Hamiltonian, as $D$ will now appear explicitly in the model.

\section{Acknowledgement}
We are thankful to the Department of Science and Technology (DST), India for financial
support.

\end{document}